\documentclass[runningheads]{llncs}

\usepackage[T1]{fontenc}
\usepackage{graphicx}
\usepackage{listings}
\usepackage{tabularx}
\usepackage{pifont}
\usepackage{xltabular}
\usepackage{multirow}
\usepackage{array}
\usepackage{float}
\usepackage{tikz}
\usepackage{booktabs}
\usepackage{xurl}
\usepackage{paralist}
\usepackage{hyperref}
\usepackage{cite}
\usepackage{xcolor}
\usepackage{pgfplots}
\usepackage{siunitx}
\usepackage{algorithm}
\usepackage{algpseudocode}
\usepackage{amsfonts}
\usepackage{enumitem,lipsum}
    
\usetikzlibrary{arrows.meta, positioning, shapes.multipart, patterns}
\pgfplotsset{compat=1.18}

\definecolor{staticPurple}{HTML}{5D2E68}
\definecolor{taintPink}{HTML}{D9706F}

\lstdefinelanguage{JavaScript}{
    keywords={break, case, catch, class, const, continue, debugger, default, delete, do, else, export, extends, finally, for, function, if, import, in, instanceof, let, new, return, super, switch, this, throw, try, typeof, var, void, while, with, yield},
    keywordstyle=\color{blue}\bfseries,
    ndkeywords={true, false, null, undefined},
    ndkeywordstyle=\color{orange}\bfseries,
    identifierstyle=\color{black},
    sensitive=true,
    comment=[l]{//},
    morecomment=[s]{/*}{*/},
    commentstyle=\color{green!50!black}\ttfamily,
    stringstyle=\color{red}\ttfamily,
}

\begin{document}
\raggedbottom

\title{Leveraging Code Slicing for LLMs-based Malicious NPM Package Detection}

\titlerunning{Code Slicing for Malicious NPM Detection}

\author{
Dang-Khoa Nguyen\inst{1}\thanks{These authors contributed equally to this work.} \and
Gia-Thang Ho\inst{1}\textsuperscript{*} \and
Quang-Minh Pham\inst{1}\textsuperscript{*} \and
Tuyet Dang-Thi Anh\inst{2} \and
Minh-Khanh Vu\inst{3} \and
Thanh Cong Nguyen\inst{4} \and
Phat T. Tran-Truong\inst{1} \and
Duc Ly Vu\inst{2}\thanks{Corresponding author.}
}

\authorrunning{Dang-Khoa Nguyen et al.}

\institute{
Ho Chi Minh City University of Technology (HCMUT), VNU-HCM, Viet Nam\\
\email{\{khoa.nguyen47245,thang.ho261104,minh.pham2212075,phatttt\}@hcmut.edu.vn}
\and
Eastern International University, Viet Nam\\
\email{\{tuyet.dangthi.cit21,ly.vu\}@eiu.edu.vn}
\and
Univ. Grenoble Alpes, CNRS, Grenoble INP, VERIMAG, UMR 5104, France\\
\email{Minh-Khanh.Vu@etu.univ-grenoble-alpes.fr}
\and
PackGuard, Viet Nam\\
\email{thanh-cong.nguyen@packguard.dev}
}

\maketitle

\begin{abstract}
Software supply chain attacks on the npm ecosystem have grown increasingly sophisticated, exploiting obfuscation and complex logic to evade detection. Large Language Models (LLMs) offer strong semantic understanding of code but face practical constraints: limited context windows and high inference costs make full-package analysis infeasible, while naive token-based splitting fragments semantic context and degrades accuracy. This paper introduces an LLM-based framework for malicious npm package detection built on code-slicing techniques. We propose an adaptation of taint-based slicing for the npm ecosystem, guided by a curated inventory of JavaScript-specific sensitive APIs, to isolate security-relevant data flows from benign boilerplate. The approach reduces the mean input token count by 99.75\% and the median by 93.7\% while preserving critical malicious behaviors. Packages relying on dynamic code generation or obfuscation yield empty slices under static analysis and require deobfuscation preprocessing, a limitation we explicitly discuss. The framework is evaluated on a dataset of more than \num{7000} malicious and benign npm packages using DeepSeek-Coder-6.7B. On the \num{2537} packages amenable to static taint analysis, taint-based slicing achieves \num{87.04}\% detection accuracy, outperforming both a naive token-splitting baseline at \num{75.41}\% and a CFG-only static slicing approach at \num{75.65}\%. These results demonstrate that semantically targeted input representations improve LLM-based detection performance beyond what is achievable through simple input-size reduction, providing an effective and computationally practical defense against evolving open-source supply-chain threats.

\keywords{Open-source software security \and Software supply chain security \and Large Language Models \and Malware Detection \and Code Slicing}
\end{abstract}

\section{Introduction}

The \texttt{npm} repository has become an increasingly attractive vector for malicious actors seeking to distribute compromised packages~\cite{zhou2025analysis}. Such attacks directly threaten millions of downstream consumers and challenge the security posture of organizational software assets~\cite{duan2020towards, scalco2022feasibility}. For example, the recent Shai-Hulud self-propagating attack~\cite{wiz_sha1_hulud_2025} involved malicious packages that spread across thousands of repositories, exfiltrated secrets, and published newly compromised payloads, thereby compromising numerous widely used packages.


To counter these threats, existing defenses span signature- or rule-based methods, machine learning (ML), and, more recently, LLMs~\cite{ohm2023sok}. Rule-based approaches integrate readily into security pipelines~\cite{vu2023bad} but are brittle against novel obfuscation~\cite{froh2023differential}. ML methods relying on hand-engineered features such as API call sequences~\cite{demirkiran2022ensemble, huang2024donapi} offer greater flexibility, yet remain labor-intensive and similarly vulnerable to feature-altering evasion~\cite{ohm2022feasibility, sejfia2022practical}.

The emergence of LLMs has introduced transformative capabilities for security tasks, including malicious code analysis and detection~\cite{al2024exploring, jelodar2025large, guo2025survey}. Recent studies demonstrate that LLM-based approaches can substantially outperform or complement conventional techniques for detecting malicious open-source packages~\cite{zahan2024leveraging, wyssevaluating}. Specifically, Zahan et al. report that LLMs achieve strong precision and recall in detecting malicious \texttt{npm} packages, surpassing traditional static analysis tools by 16\% and 9\% in precision and F1 score, respectively~\cite{zahan2024leveraging}. This advantage stems from the capacity of LLMs for semantic reasoning and context-aware pattern recognition across complex code structures~\cite{kojima2022large, wei2022chain}. 
However, LLMs face a practical constraint: context windows ranging from 4K to 16K tokens are often insufficient for full-package analysis~\cite{kwon2023efficient, liang2022holistic, wyssevaluating}. When malicious logic is scattered across a codebase, the model lacks the global visibility needed for accurate threat assessment, degrading detection performance~\cite{guo2024deepseek, wyssevaluating}. To mitigate this bottleneck, Wang et al.~\cite{wang2025ocsbert} proposed a taint-based code-slicing technique that isolates security-relevant code segments before the segments are fed to LLMs. Their approach reduces context-window overhead while preserving critical semantic information by generating \textit{Sensitive Code Slices (SCS)} through taint analysis. This method achieves substantial gains in detection precision and computational efficiency for PyPI packages. However, the applicability of this slicing-based methodology to npm packages remains unexplored.


The npm ecosystem poses distinct challenges because packages typically contain numerous files, external dependencies, and utility functions that can bury malicious payloads within benign boilerplate~\cite{ohm2023sok}. This contextual noise dilutes semantic signals and exacerbates context-window constraints~\cite{kwon2023efficient, guo2024deepseek}. The study therefore examines how package context can be reduced while preserving semantic signals of malicious behavior, and how these reduced representations affect LLM-based package classification. Our  evaluation compares taint-based slicing using data-flow graphs, static slicing using control-flow graphs, and naive token splitting as package representations for DeepSeek-Coder-6.7B. Context reduction is measured in terms of lines of code and token counts, while the preservation of security-relevant information is measured using Sensitive Feature Recall (SFR). Classification performance is then evaluated on a curated dataset of malicious and benign npm packages.

In this paper, we make the following three contributions.
\begin{itemize}
\item First, we adapt and evaluate two code-slicing strategies for npm packages using Joern Code Property Graph analysis and a curated JavaScript-specific API taxonomy.
\item Second, we quantify the reduction in context size and the preservation of security-relevant information produced by these slicing strategies.
\item Third, we evaluate how sliced representations affect LLM-based malicious package detection compared with a naive splitting baseline.
\end{itemize}

\section{Background}

\label{sec:background}

This section reviews three lines of work that motivate the proposed approach: syntactic and signature-based detection, behavioral and neural malware detection, and LLM-oriented code analysis. It also introduces the program representations and slicing concepts underlying the proposed methodology.

\subsection{Syntactic Pattern Matching and Signature-Based Detection}

Early malware detection relied primarily on lexical and syntactic pattern matching. Tools such as Microsoft OSS Detect Backdoor (ODB)~\cite{microsoft2023applicationinspector, microsoft2023ossdetectbackdoor} perform registry and file-system pattern scanning, whereas CodeQL-based systems~\cite{froh2023differential} use rule-based syntax-tree traversal to identify known threats. Although these approaches offer linear scalability and low detection latency, they exhibit a notable precision and recall trade-off. For example, Microsoft's ODB reportedly reaches a 75.5\% false-positive rate~\cite{microsoft2023applicationinspector}, which is impractical for maintainers who require very low false-positive rates~\cite{vu2023bad}. Rule-based systems are also brittle against obfuscation techniques (e.g., control-flow flattening)~\cite{froh2023differential} and suffer from signature lag, creating vulnerability windows for polymorphic variants~\cite{duan2020towards}. To mitigate these limitations, prior work has explored metadata-augmented signals (e.g., package version history and developer reputation)~\cite{duan2020towards} and discrepancy analysis between source code and published package artifacts~\cite{vu2021lastpymile}.

\subsubsection{Behavioral sequence analysis through API monitoring}\label{sec:behavioral-sequence-analysis}

Existing studies used Recurrent Neural Networks (RNNs)~\cite{pascanu2015malware} to model API call sequences, but their accuracy was limited by vanishing/exploding gradients and data scarcity. Subsequent hybrid models, such as CNN-RNN ensembles~\cite{demirkiran2022ensemble} and TextCNN~\cite{kim2014convolutional}, improved performance by capturing both local and sequential patterns. However, dynamic analysis still faces two critical hurdles. First, behavioral sequences often lack explicit semantic intent; Yan et al.~\cite{yan2023prompt} reported weak semantic cohesion in API associations. Second, malware increasingly employs anti-analysis techniques (e.g., environment-adaptive execution) to evade dynamic approaches~\cite{li2021detecting}. Moreover, sandboxed execution incurs computational costs proportional to package complexity, which precludes its use in real-time registry monitoring contexts~\cite{huang2024donapi}.

\subsection{Semantic Representation Learning via Feature Engineering and Neural Methods}

To address the limited semantic depth of dynamic analysis, recent research has shifted toward learning representations directly from code. Initial efforts relied on either extracted code features combined with support vector machine (SVM) classifiers~\cite{sejfia2022practical} or handcrafted metrics~\cite{ohm2022feasibility}. While these approaches can be effective for specific threat classes, they depend on labor-intensive, subjective feature engineering, which undermines reproducibility~\cite{ohm2022feasibility} and reduces robustness against semantic-preserving obfuscations~\cite{froh2023differential}. Scope is also a limitation: DONAPI~\cite{huang2024donapi} restricts static analysis to entry points (e.g., installation scripts), risking oversight of dormant payloads in nested files, while install-time trigger methods~\cite{sejfia2022practical} struggle against runtime logic bombs and novel obfuscation.

\subsection{Large Language Models for Semantic Code Analysis}

Recent advances in Large Language Models (LLMs) have shifted the analytical paradigm toward higher-level semantic understanding. Encoder-only models such as CodeBERT~\cite{feng2020codebert} and GraphCodeBERT~\cite{guo2020graphcodebert} leverage bimodal pretraining and Program Dependence Graphs (PDGs), but their context windows of 512 to 1024 tokens limit full-package analysis~\cite{devlin2019bert}. Generative LLMs (e.g., GPT-4 and DeepSeek-Coder) go further by reasoning about higher-order program intent~\cite{wei2022chain}, improving robustness against intent-preserving obfuscation~\cite{zahan2024leveraging} and supporting cross-language transfer without explicit retraining~\cite{guo2024deepseek}.

When applying LLMs to malicious npm package detection, critical architectural constraints remain. First, the context windows of many standard models (typically 4K to 16K tokens) are often smaller than the size of many complete packages~\cite{kwon2023efficient, liang2022holistic}. Truncating code to fit these context-window constraints can remove critical semantics and reduce detection accuracy, especially when malicious payloads are distributed across multiple files~\cite{guo2024deepseek, zhu2024deepseek}. Second, high memory demands and inference latency make continuous, real-time scanning of production registries computationally expensive~\cite{balan2025}.

Although extended-context LLMs (some supporting 128K or more tokens) reduce context-window pressure, slicing remains necessary for two reasons. First, benign scaffolding acts as semantic noise even within long windows; taint-based slicing retains only security-critical data flows, improving precision regardless of context capacity. Second, the largest packages in our dataset reach 15.8 million tokens, which is far beyond any current model. Full-scale processing would incur prohibitive memory and latency for real-time registry monitoring. Slicing reduces the average input to~403 tokens, with some packages reduced by as much as a factor of 237.

\subsection{Graph Representations of Programs}

Graph-based program representations capture semantics, control flow, and data dependencies that are obscured in raw text~\cite{Johnson1993Dependence, Callahan1988ProgramSummary}. An Abstract Syntax Tree (AST) encodes syntactic structure hierarchically and serves as the foundation for richer representations. A Control Flow Graph (CFG) is a directed graph $G_{CFG}=(V,E)$ where nodes are basic blocks and edges are control-flow transitions, thereby capturing feasible execution paths. A Data Dependence Graph (DDG) records def-use relationships: an edge $(u,v)$ exists when statement $u$ defines a variable whose value reaches statement $v$ without intervening reassignment~\cite{Hammer2009FlowSensitive}, making it the foundation for code slicing. Finally, a Code Property Graph (CPG)~\cite{Yamaguchi2014ModelingAD} unifies the AST, CFG, and DDG into a single property multigraph, enabling cross-layer security queries. Our approach uses CPGs (via Joern~\cite{joern}) to extract security-relevant slices from npm package source code.

\subsection{Code Slicing Techniques}

\paragraph{Static Code Slicing} computes the transitive closure of data and control dependencies from a slicing criterion $C = (s, v)$ by traversing a Program Dependence Graph (PDG)~\cite{Ottenstein1984ThePD}, yielding a reduced executable that preserves the behavior of the original program $P$ for all possible inputs~\cite{Tip1994ASO}. Because it assumes no knowledge of runtime values, it conservatively includes all statements that could affect the criterion, which tends to produce larger slices compared with dynamic approaches.

\paragraph{Taint-based Slicing} focuses on security-critical data flows by tracking untrusted data from a \textit{Source} (e.g., user inputs, network payloads) through assignments and computations to a \textit{Sink} (e.g., command execution, file-system access)~\cite{Krinke2007InformationFlow, Wilander2005Modeling}. These propagation paths are mapped via Data Dependence Graphs (DDGs) or Code Property Graphs (CPGs)~\cite{Halim2019StaticAnalyzer, Li2020DepTaint}, isolating the precise attack chain from surrounding benign code.
\subsection{Comparative Related Work}

\label{sec:related_work}

Ahmed et al.~\cite{ahmed2024automatic} abstract code into 16 high-level behavioral actions and analyze their chronological order, whereas our taint-based slicing traces how sensitive data flows through the code, removing irrelevant boilerplate while preserving security-relevant detection signals. Huang et al.~\cite{huang2024spiderscan} detect malicious npm packages via graph-based behavior matching, which requires a predefined attack template for each threat class. Our approach feeds raw code slices directly to an LLM, enabling template-free semantic reasoning over novel attack patterns.

Wang et al.~\cite{wang2025malpacdetector} split package source code into naïve fixed-size segments and instruct GPT-3 to summarize malicious behaviors. Our taint-based slicing achieves greater token reduction and higher detection accuracy on the same packages (Sections~\ref{sec:rq1} and~\ref{sec:rq2}). Zahan et al.~\cite{zahan2024leveraging} apply domain-specific zero-shot prompting to npm malware detection. Our contribution is orthogonal in that we selectively isolate security-relevant code fragments before prompting, improving both input quality and LLM reasoning precision.

Our work is most closely related to OCS-BERT~\cite{wang2025ocsbert}, which applies taint-based slicing to PyPI malware detection using a discriminative BERT model. We extend this paradigm to the npm ecosystem in two key ways. First, it targets generative LLMs for direct semantic security scoring rather than training classifiers on slice embeddings. Second, it systematically compares BFS-based static CFG slicing with a refined taint-based slicing approach using Joern's \texttt{ReachableByFlows} API, explicitly quantifying the noise reduction and detection performance trade-offs of each code slicing strategy.

\section{Methodology}
\label{sec:methodology}

\begin{figure}[!ht]
    \centering
    \includegraphics[width=1\textwidth]{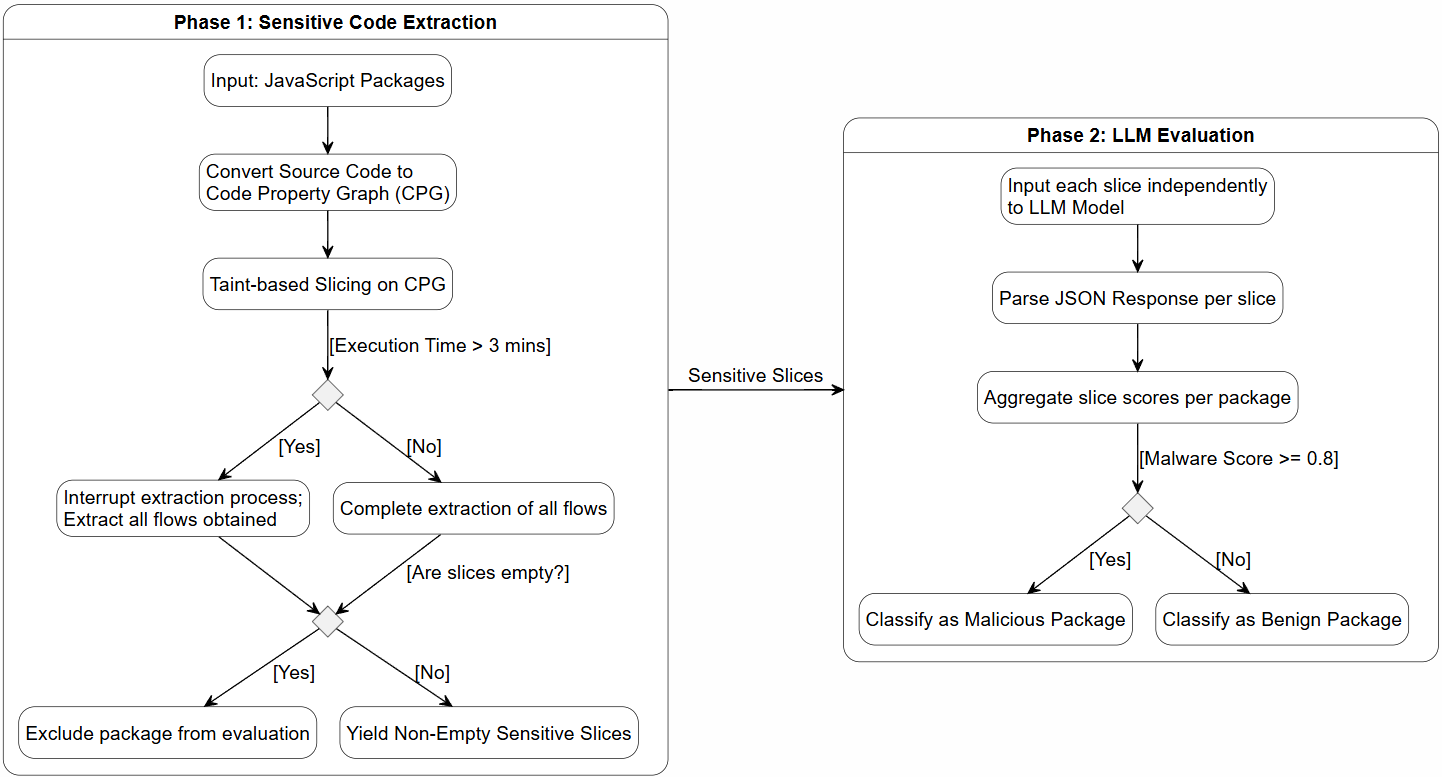}
    \caption{Our NPM malware detection workflow}
    \label{fig:workflow}
\end{figure}

Our methodology, illustrated in Figure~\ref{fig:workflow}, presents a slicing-based malware-detection pipeline for npm packages. By extracting security-relevant code slices and feeding them into a large language model, we infer potential malicious intent from localized semantic patterns. The slice-level predictions produced by the LLM are then aggregated to yield a package-level malware assessment that determines whether the package is malicious.

\subsection{Sensitive Code Extraction} \label{extraction_step}
In the first step, we isolate security-relevant code slices from npm packages
using static and taint-based slicing strategies. Our goal is to assess how the different
code representations produced by these techniques affect the signal-to-noise ratio in malware
detection, particularly within the npm ecosystem.

The static code slicing technique shown in  Algorithm~\ref{alg:static_slicing} performs a backward breadth-first search (BFS) traversal on the CFG from sensitive sink nodes, collecting all predecessor statements that structurally influence the path to the sink. While this captures broad control dependencies, it does not track data propagation and therefore tends to produce larger, noisier slices~\cite{prasanna2019static}.

\algtext*{EndWhile}
\algtext*{EndIf}
\algtext*{EndFor}
\algtext*{EndProcedure}
\algtext*{EndFunction}

\setlength{\intextsep}{3pt}
\setlength{\textfloatsep}{4pt plus 1pt minus 1pt}
\setlength{\floatsep}{4pt plus 1pt minus 1pt}
\setlength{\abovecaptionskip}{2pt}
\setlength{\belowcaptionskip}{2pt}

\begin{algorithm}[H]
\caption{Control-Flow-Based Static Slicing}
\label{alg:static_slicing}
\footnotesize
\begin{algorithmic}[1]
\Procedure{StaticSlicing}{$cpg, sourceGroup, sinkGroup$}
    \State $sourceIds \gets \emptyset$, $slices \gets \emptyset$
    \ForAll{$source \in sourceGroup$}
        \State $sourceIds \gets sourceIds \cup \Call{QueryIds}{cpg, source}$
    \EndFor
    \ForAll{$sink \in \Call{Query}{sinkGroup}$}
        \State $slice \gets \Call{ReachableByFlow}{sink, sourceIds}$
        \If{$slice \neq \emptyset$}
            \State $slices \gets slices \cup \{\Call{FilterClean}{slice}\}$
        \EndIf
    \EndFor
    \State \Return $slices$
\EndProcedure
\Statex
\Function{ReachableByFlow}{$sink, sourceIds$}
    \State $worklist \gets [sink]$, $visited \gets \{sink\}$
    \State $slice \gets \emptyset$, $sourceReached \gets \textsc{false}$
    \While{$worklist \neq \emptyset$}
        \State $node \gets$ remove next element from $worklist$
        \State $slice \gets slice \cup \{node\}$
        \If{$node.id \in sourceIds$}
            \State $sourceReached \gets \textsc{true}$
        \EndIf
        \ForAll{$pred \in node.\mathrm{in}(\mathrm{CFG})$}
            \If{$pred \notin visited$}
                \State $visited \gets visited \cup \{pred\}$
                \State add $pred$ to $worklist$
            \EndIf
        \EndFor
    \EndWhile
    \If{$sourceReached$}
        \State \Return $slice$
    \Else
        \State \Return $\emptyset$
    \EndIf
\EndFunction
\end{algorithmic}
\end{algorithm}

The Taint-based slicing shown in Algorithm~\ref{alg:taintBasedSlicing} uses Joern's \texttt{ReachableByFlows} API~\cite{Yamaguchi2014ModelingAD} to trace data-flow paths from untrusted sources to critical sinks. Our contribution lies in the JavaScript-specific Source–Sink–Dual taxonomy (Table~\ref{tab:feature_groups}), which guides traversal and reduces contextual noise before the slices are input to the LLM. The API inventory was built by adapting the PyPI-centric taxonomy of Wang et al.~\cite{wang2025ocsbert} to Node.js (e.g., replacing \texttt{subprocess} with \texttt{child\_process.exec}) and seeding it with JavaScript-specific functions from Yu et al.~\cite{yu2024maltracker}. We then augmented the list by analysing publicly disclosed npm attacks from Snyk~\cite{snyk_advisories}, Phylum~\cite{phylum_blog}, Checkmarx~\cite{checkmarx_blog}, and the MalPacDetector dataset~\cite{wang2025malpacdetector}, adding any APIs used for data exfiltration, payload execution, or obfuscation. Dynamic-generation APIs (e.g., \texttt{eval}, \texttt{Function()}, \texttt{Buffer.from(\ldots,\,`base64')}) are explicitly designated as Sinks so that the slicer preserves data-flow paths into obfuscated executions. The resulting inventory (Table~\ref{tab:feature_groups}) assigns each API one of three roles: \textbf{Source} (e.g., \texttt{process.env}), \textbf{Sink} (e.g., \texttt{exec}, \texttt{eval}), or \textbf{Dual} (e.g., \texttt{fs.readFileSync}/\texttt{fs.writeFileSync}).

\begin{algorithm}[t]
\caption{Taint-Based Slicing Procedure}
\label{alg:taintBasedSlicing}
\begin{algorithmic}[1]

\Procedure{TaintBasedSlicing}{$sourceGroup, sinkGroup$}

    \State $slicedPaths \gets \emptyset$

    \ForAll{$sourceDef \in sourceGroup$}

        \State $sourceNodes \gets \Call{Query}{sourceDef}$

        \If{$sourceNodes \neq \emptyset$}

            \ForAll{$sinkDef \in sinkGroup$}

                \State $sinkNodes \gets \Call{Query}{sinkDef}$

                \If{$sinkNodes \neq \emptyset$}

                    \State $flows \gets
                    \Call{ReachableByFlows}{sourceNodes, sinkNodes}$

                    \If{$flows \neq \emptyset$}
                        \State $slicedPaths \gets slicedPaths \cup flows$
                    \EndIf

                \EndIf

            \EndFor

        \EndIf

    \EndFor

    \State \Return $slicedPaths$

\EndProcedure

\end{algorithmic}
\end{algorithm}

The categorized APIs are used to guide the slicing process toward security-relevant code paths. To ensure scalability and prevent excessively large slices, we enforce a strict timeout of three minutes per package. This constraint is particularly important for complex npm packages, where large codebases can produce expansive graphs. To determine this timeout, multiple experiments were conducted to measure the success rate of the slicing techniques across a large set of npm packages. We observed that most packages completed slicing within the allotted timeframe, while only a small fraction exceeded the limit.

\begin{table}[H]
\small
\centering
\caption{Categorization of Sensitive APIs and Their Security Roles}
\label{tab:feature_groups}
\renewcommand{\arraystretch}{1.2}
\setlength{\tabcolsep}{4pt}
\begin{tabularx}{\linewidth}{@{}l l >{\raggedright\arraybackslash}X c@{}}
\hline
\textbf{Role} & \textbf{Feature Group} & \textbf{Sensitive JavaScript APIs} & \textbf{\#APIs} \\ \hline
Source & Information Gathering & \path{process.env}, \path{os.userInfo} & 57 \\ \hline
\multirow{4}{*}{Sink} & System Execution & \path{child_process.exec}, \path{spawn}, \path{eval} & 21 \\ \cline{2-4}
& Code Obfuscation & \path{Buffer.from(...,'base64')}, \path{JSON.stringify} & 19 \\ \cline{2-4}
& Environment Cleanup & \path{fs.unlink}, \path{process.kill} & 29 \\ \cline{2-4}
& Parallel Processing & \path{cluster.fork} & 2 \\ \hline
\multirow{2}{*}{Dual} & File Operations & \textit{Source:} \path{fs.readFileSync}; \textit{Sink:} \path{fs.writeFileSync} & 85 \\ \cline{2-4}
& Network Communication & \textit{Source:} \path{socket.on}, \path{createServer}; \textit{Sink:} \path{http.post} & 43 \\ \hline
\end{tabularx}
\end{table}

Next, we perform an automated security scoring with LLMs.
The slices from Step~\ref{extraction_step} are fed into the selected LLM (see Subsection~\ref{sec:model_criteria} for selection criteria) for automated malicious-code assessment. The slices of each package are submitted to the LLM independently, and the model returns a JSON-formatted response per slice.
The system prompt was adapted from the design strategies of Yu et al.~\cite{yu2024maltracker} and tailored to DeepSeek-Coder-6.7B through iterative validation on pilot outputs. Three design decisions were made: (1) a \emph{static analysis persona} that prevents the model from simulating execution; (2) a scoring rubric that defines four output fields, namely \texttt{confidence}, \texttt{obfuscated}, \texttt{malware}, and \texttt{securityRisk}, each taking values in $S \in [0,1]$ with explicit thresholds; and (3) a \emph{strict JSON output schema} with enforced key ordering that prohibits surrounding markdown, enabling reliable parsing by the aggregation stage. The complete prompt is available in the replication package. The \texttt{malware} score of each slice is extracted and passed to aggregation and package evaluation phase.

In the final stage, we aggregate the responses returned by the LLM to assess the overall maliciousness of an npm package. Because a package may yield multiple responses corresponding to individual code snippets, we adopt a conservative aggregation strategy: the maximum maliciousness score across all snippets is used as the package-level indicator. The LLM processes each snippet and returns a JSON object containing a maliciousness score \( S \in [0,1] \).
We selected the detection threshold  $\tau = 0.8$, following the highest-risk designation of Zahan et al.~\cite{zahan2022weak}. Because the max-aggregation strategy flags a package if \emph{any} single slice exceeds $\tau$, the threshold must counterbalance the resulting sensitivity to isolated high-risk patterns: a high $\tau$ requires near-certain model confidence, reducing false positives from benign packages that legitimately invoke sensitive APIs (e.g., \texttt{child\_process} in build tools). Empirical sensitivity analysis is reported in Section~\ref{sec:rq2}. A package is classified as malicious if $S \geq \tau$, and benign otherwise.

\section{Evaluation}
\label{sec:experiment-setup}

The evaluation examines how different slicing strategies reduce contextual noise while preserving malicious patterns, and how these representations affect LLM detection performance. Three code representations are evaluated. The naive splitting baseline uses full source code divided into non-overlapping \num{500}-token segments without slicing. It was re-implemented locally based on Wang et al.~\cite{wang2025malpacdetector} and evaluated on the same \num{2537}-package subset with the same model, prompt, aggregation operator, and threshold ($\tau = 0.8$) as the slicing strategies. Static slicing uses CFG-based backward reachability to capture broad control dependencies without tracking data flow. Taint-based slicing integrates CFG and DFG information to track data flow from sources to sensitive sinks (Section~\ref{sec:methodology}), focusing the LLM on code regions where malicious data is actively manipulated.

We implemented the slicing algorithms using the Joern framework~\cite{Yamaguchi2014ModelingAD} (JRE 21, Scala 2.13, sbt 1.11.7) to leverage its code property graph (CPG) capabilities for JavaScript analysis. The experiments utilized the \texttt{DeepSeek-Coder-6.7B} model, deployed using model parallelism with half-precision (FP16) weights on a server equipped with an Intel Core i9-10900X CPU, \num{64}~GB of RAM, and two NVIDIA GeForce RTX 3080 GPUs (each with 10~GB of VRAM).

\subsection{Data Preparation}
\label{sec:dataset}

We utilize the MalnpmDB dataset~\cite{wang2025malpacdetector}, comprising \num{4051} benign and \num{3258} malicious npm packages (\num{7309} total) collected from real-world repositories and independently verified by the dataset authors. As the dataset is not publicly available, we contacted the authors to obtain access. We applied a three-stage preprocessing pipeline summarised in Table~\ref{tab:dataset_pipeline}. In the first stage, we remove samples with corrupt or empty archives. In the second stage, we exclude metadata-only packages and those invoking no sensitive APIs (Table~\ref{tab:feature_groups}), yielding \num{3242} benign and \num{2600} malicious packages. In the final stage, we apply the slicing pipeline with a three-minute per-package timeout; packages producing no non-empty slices were excluded (root causes discussed in Section~\ref{sec:threats-to-validity}). This resulted in non-empty slices for \num{1084} benign ($\approx$33\%) and \num{1453} malicious ($\approx$56\%) packages, each contributing multiple individual slices.

\begin{table}[H]
    \footnotesize
    \centering
    \caption{Package counts after preprocessing}
    \label{tab:dataset_pipeline}
    \setlength{\tabcolsep}{4pt}
    \begin{tabular}{lccc}
        \toprule
        \textbf{Stage} & \textbf{Benign} & \textbf{Malicious} & \textbf{Total} \\
        \midrule
        Raw MalnpmDB & \num{4051} & \num{3258} & \num{7309} \\
        After Stage 1 (archive validation) & \num{3242} & \num{2731} & \num{5973} \\
        After Stage 2 (content filtering) & \num{3242} & \num{2600} & \num{5842} \\
        After Stage 3 (non-empty slices) & \num{1084} (33\%) & \num{1453} (56\%) & \num{2537} \\
        \bottomrule
    \end{tabular}
\end{table}

\subsection{Large Language Model Selection}
\label{sec:model_criteria}

LLMs offer varying capabilities and constraints. Selecting an LLM for malicious-code detection requires balancing analytical performance against resource limitations. Our model selection is guided by criteria inspired by prior work~\cite{liang2022holistic, chen2021evaluating} and constrained by our computing environment.

The selected model must reason about data flow, control flow, and API usage while producing strictly parseable JSON responses. Encoder-only models such as BERT~\cite{devlin2019bert} were excluded because they lack generative capability. CodeLlama~\cite{roziere2023code} and StarCoder~\cite{li2023starcoder} were excluded after pilot trials on 100 held-out slices showed frequent non-compliance with the strict JSON-only output schema, including markdown fences and incomplete objects. DeepSeek-Coder's \num{16}K-token window covers approximately 90\% of slices without truncation ($\mu=\num{403}$, max $=\num{28979}$ tokens)~\cite{guo2024deepseek}, and it achieved near-perfect adherence to the JSON output schema in pilot trials. The model is also MIT-licensed and deployable without quantization on two RTX 3080 GPUs (20~GB total VRAM), supporting reproducibility and eliminating API costs.

\newcommand{\xmark}{\ding{55}}

\begin{table}[H]
    \centering
    \caption{Comparison of Candidate LLMs for Malicious Package Detection}
    \label{tab:table_model_comparison}
    \small
    \setlength{\tabcolsep}{4pt}
    \renewcommand{\arraystretch}{1.15}
    \resizebox{\linewidth}{!}{%
    \begin{tabular}{@{}lllllll@{}}
        \toprule
        \textbf{Model} & \textbf{Params.} & \begin{tabular}{@{}l@{}}\textbf{Context} \\ \textbf{Length}\end{tabular} &
        \textbf{License} & \textbf{VRAM} &
        \begin{tabular}{@{}c@{}}\textbf{Code} \\ \textbf{Understanding}\end{tabular} &
        \begin{tabular}{@{}c@{}}\textbf{JSON} \\ \textbf{Adherence}\end{tabular} \\
        \midrule
        GPT family~\cite{openai2024gpt4technicalreport} & Unknown & 8K--128K+ & Commercial & N/A & High & High \\
        BERT~\cite{devlin2019bert} & 125M--400M & 512--1024 & Open source & $\sim$0.5--1GB & Low & N/A \\
        CodeLlama~\cite{roziere2023code} & 7B--70B & 4K--32K & Open source & 14--40GB+ & Medium & Low \\
        StarCoder~\cite{li2023starcoder} & 3B--21B & 8K--16K & Open source & 6--24GB+ & Medium--low & Low \\
        \textbf{DeepSeek-Coder}~\cite{guo2024deepseek} & \textbf{1.3B--33B} & \textbf{16K} & \textbf{Open source} & \textbf{2.6--66GB} & \textbf{Medium--high} & \textbf{High} \\
        \bottomrule
    \end{tabular}%
    }
\end{table}

Table~\ref{tab:table_model_comparison} summarizes the candidate LLMs considered in this study. We selected the \num{6.7}B instruction-tuned DeepSeek-Coder model~\cite{guo2024deepseek}, which combines strong code-understanding capability (trained on $\sim$2T tokens, of which 87\% are code) with a favorable trade-off between efficiency and effectiveness, suited to our resource-constrained environment.

\section{Context Reduction with Code Slicing} \label{sec:context-reduction}\label{sec:rq1}

We evaluate both slicing approaches across three dimensions: context reduction, preservation of malicious context, and preservation of semantic integrity, measured relative to the naive splitting baseline.

\subsection{Context Reduction Rate}

Table~\ref{tab:size_stats} reports the statistical distribution of context size for all three methods, measured in both number of Lines of Code (LoC) and subword tokens. Among the evaluated techniques, the naive splitting baseline exhibits extreme variability: \num{1} to \num{544680} LoC and \num{69} to \num{15765669} tokens, with mean–median divergence (\num{8002} vs.\ \num{306} LoC; \num{162920} vs.\ \num{3594} tokens) indicating a highly skewed distribution driven by a small number of disproportionately large packages.

\begin{table}[H]
    \centering
    \footnotesize
    \caption{Statistical distribution of context size (LoC and tokens).}
    \label{tab:size_stats}
    \setlength{\tabcolsep}{3pt}
    \renewcommand{\arraystretch}{1.0}
    \resizebox{\columnwidth}{!}{%
    \begin{tabular}{llrrrrrr}
        \toprule
        \textbf{Unit} & \textbf{Method} & \textbf{Mean} & \textbf{Min} & \textbf{Q25} & \textbf{Median} & \textbf{Q75} & \textbf{Max} \\
        \midrule
        \multirow{3}{*}{LoC}
          & Naive Splitting    & \num{8002} & \num{1} & \num{41}  & \num{306} & \num{3109}  & \num{544680} \\
          & Static Slicing     & \num{32}   & \num{1} & \num{11}  & \num{31}  & \num{38}    & \num{1122}   \\
          & Taint-based Slicing & \num{32}  & \num{1} & \num{4}   & \num{18}  & \num{28}    & \num{2247}   \\
        \midrule
        \multirow{3}{*}{Tokens}
          & Naive Splitting    & \num{162920} & \num{69} & \num{408} & \num{3594} & \num{40219} & \num{15765669} \\
          & Static Slicing     & \num{379}    & \num{9}  & \num{158} & \num{358}  & \num{431}   & \num{13596}    \\
          & Taint-based Slicing & \num{403}   & \num{3}  & \num{76}  & \num{228}  & \num{307}   & \num{28979}    \\
        \bottomrule
    \end{tabular}%
    }
\end{table}


Both slicing methods substantially reduce and normalize context sizes (Table~\ref{tab:size_stats}). Median LoC drops to \num{31} (static slicing) and \num{18} (taint-based slicing), and the maximum shrinks from \num{545}k to under \num{2300} LoC, representing a \num{237}$\times$ reduction. On the token side, mean input falls from \num{162920} to \num{379} (static) and \num{403} (taint-based), while the median decreases from \num{3594} to \num{228}, yielding reductions of \textbf{99.75\%} and \textbf{93.7\%}, respectively. These results confirm that these gains apply across the full distribution rather than being driven solely by extreme outliers.

\subsection{Preservation of Malicious Context}


Context reduction is counterproductive if it discards malware-relevant semantics. We quantify retention using \emph{Sensitive Feature Recall} (SFR), defined as the fraction of sensitive API references (sources and sinks) preserved after slicing:
\begin{equation}
    SFR = \frac{|F_{sliced} \cap F_{original}|}{|F_{original}|} \times 100\%,
\end{equation}
where \(F_{original}\) is the set of source-code lines referencing a sensitive API (e.g., \texttt{process.env} as a source, \texttt{exec} as a sink) in the original package, and \(F_{sliced}\) is the subset retained in the slice.

By analyzing slices generated by static slicing and taint-based slicing, we observed a clear contrast in how each method handles benign and malicious code. In benign npm packages, static slicing retains approximately \num{15.9}\% of sensitive API features, whereas taint-based slicing is more selective, retaining only \num{9.4}\%. Conversely, for malicious packages, both methods exhibit high retention rates, with taint-based slicing achieving a slightly higher SFR (\num{73.7}\%) than static slicing (\num{73.4}\%). This result indicates that taint-based slicing is more effective at filtering incidental sensitive API calls (i.e., noise) in benign applications without compromising preservation of malicious information flows. In practical deployment, this behavior can reduce false positives, a high priority for package-repository maintainers~\cite{vu2023bad}.

Overall, taint-based slicing reduces the mean token count by 99.75\% and the median token count by 93.7\%. It retains 73.7\% of sensitive API features in malicious packages but only 9.4\% in benign packages, indicating that the compression primarily removes irrelevant contextual noise.

\section{Detection Performance on Code Slices}\label{sec:detection-performance}\label{sec:rq2}

Table~\ref{tab:result_stats} compares DeepSeek-Coder-6.7B performance across the three input representations on the identical \num{2537}-package subset. Both slicing techniques outperform naive splitting across all metrics. Taint-based slicing achieves the strongest gains: +\num{11.39}\% accuracy and +\num{9.11}\% F1-score over naive splitting, and +\num{11.39}\% accuracy over static slicing. Because the evaluation set is held constant, these gains are attributable solely to input representation quality, not to sample selection differences.

The figures in Table~\ref{tab:result_stats} apply only to the \num{2537} packages that yielded non-empty slice sets (56\% of malicious, 33\% of benign); the remaining 44\% of malicious and 67\% of benign packages were not captured by the slicer due to dynamic runtime API construction patterns. This \emph{survival bias} implies that the reported 87.04\% accuracy overestimates performance on the full corpus. In production, packages producing empty slices must be routed to either dynamic sandboxing or LLM-assisted deobfuscation before re-entering the pipeline (see Section~\ref{sec:threats-to-validity}).

\begin{table}[!t]
    \footnotesize
    \begin{center}
        \caption{Detection performance by input representation}
        \label{tab:result_stats}
        \setlength{\tabcolsep}{3pt}
        \begin{tabular*}{\columnwidth}{@{\extracolsep{\fill}}lllll@{}}
            \toprule
            \textbf{Slicing Technique} & \textbf{Accuracy} & \textbf{Precision} & \textbf{Recall} & \textbf{F1 Score} \\
            \midrule
            Static Slicing & \num{75.65} & \num{95.98} & \num{60.49} & \num{74.21} \\
            Taint-based Slicing & \textbf{\num{87.04}} & \textbf{\num{96.23}} & \textbf{\num{72.90}} & \textbf{\num{83.32}} \\
            Naive Splitting & \num{75.41} & \num{87.05} & \num{55.52} & \num{67.80} \\
            \bottomrule
        \end{tabular*}%
    \end{center}
\end{table}


Naive splitting disrupts interprocedural dependencies across fixed-size chunks, making program intent harder for the LLM to infer. Static slicing restores control-flow reachability and improves F1 to \num{74.21}\%, but its broader scope retains benign scaffolding that dilutes the malicious signal in the input, lowering recall to \num{60.49}\%. Taint-based slicing narrows analysis to paths influenced by untrusted sources. It raises recall to \num{72.90}\% (up from \num{60.49}\% for static slicing) while preserving similarly high precision (\num{96.23}\% vs. \num{95.98}\%).

To validate the choice of $\tau = 0.8$, we evaluated taint-based slicing across $\tau \in {0.5, 0.7, 0.8, 0.9}$. Lower thresholds substantially reduce precision (59.81\% at $\tau=0.5$; 64.35\% at $\tau=0.7$) because benign packages with unusual but safe API usage patterns produce moderately elevated snippet scores that exceed the decision boundary. Raising the threshold to 0.9 restores precision but further reduces recall below 72.90\%, increasing the rate of false negatives. $\tau=0.8$ achieves the best F1 score (83.32\%) and aligns with the highest-risk threshold defined by Zahan et al.~\cite{zahan2022weak}. The score distribution is strongly bimodal: benign snippets cluster near $S \approx 0.1$ and malicious snippets near $S \approx 0.9$, so adjusting the threshold within that range affects only a small set of ambiguous cases.


\section{Threats to Validity}
\label{sec:threats-to-validity}

\paragraph{Malicious payloads may reside in metadata files (e.g., \texttt{package.json}).} However, samples containing only this file were excluded to isolate and evaluate the effectiveness of the proposed slicing techniques for \emph{JavaScript source-code logic} rather than metadata-based attack vectors.

\paragraph{Exclusion of packages with empty slices due to processing timeouts.} Packages that yielded empty slices were excluded due to strict processing timeouts. Hardware constraints also occasionally led to memory errors during inference on extremely large slices. Although these issues may affect accuracy for a small number of outlier cases, the high overall recall (\num{96.23}\%) observed in the results suggests that such edge cases do not materially impact the overall validity of the proposed approach.

\paragraph{Information loss in code-slicing techniques.} Removing code segments may eliminate contextual cues necessary for effective LLM-based reasoning. However, an empirical analysis using Sensitive Feature Recall (SFR) helps mitigate this concern. The results show that taint-based slicing retains more than \num{73}\% of malicious API features while discarding almost \num{90}\% of benign code segments. Moreover, the substantial improvement in F1-score (\num{83.32}\%) compared to the naive splitting baseline (\num{67.80}\%) indicates that the sliced representation enhances semantic signal quality rather than degrading it.

\paragraph{Sensitive Feature Recall (SFR) as a lexical proxy rather than a semantic metric.} SFR measures the reachability of known API markers but cannot capture complex malicious patterns such as dynamic string construction, indirect invocations, or obfuscated encoding chains. JavaScript’s pervasive \texttt{async}/\texttt{await} constructs and callback patterns frequently break static data-flow edges, causing semantically linked source-to-sink pairs to appear disconnected in the CPG representation.

\paragraph{Vulnerability to code obfuscation and dynamic code generation.} Dynamic constructs such as \texttt{eval}, \texttt{Function()}, and runtime string construction hide dynamically resolved call targets from static CFG/DFG analysis. This limitation produces empty slices and helps explain the 44\% empty-slice rate for malicious packages. The reported accuracy in this paper therefore reflects a survival-biased subset; deobfuscation preprocessing is required before the pipeline can be applied to heavily obfuscated packages.

\enlargethispage{\baselineskip}

\paragraph{Limitation to a single open-source LLM family in evaluation.} This improves reproducibility and eliminates API costs, but leaves performance on proprietary models (e.g., GPT-4 and similar proprietary models) unexplored. The core benefit of taint-based slicing—namely higher input information density—is expected to be model-agnostic. Future work should validate this claim on closed-source models and help establish an upper bound on achievable performance.

\paragraph{Bias toward predefined sensitive API coverage.} Following Yu et al.~\cite{yu2024maltracker}, we curated our API inventory from known attack patterns in prior malware datasets~\cite{arp2014drebin,duan2020towards}, rendering the framework unable to detect previously unseen vectors that exploit novel Node.js primitives or previously unseen indirect call chains absent from the inventory. Future work should replace static curation with automated dynamic API discovery mechanisms~\cite{gao2025malguard} to keep source/sink definitions current without manual effort.

\paragraph{Potential training data contamination and inflation of absolute metrics.} Popular benign npm libraries were likely included in DeepSeek-Coder’s pre-training corpus, and some publicly disclosed malicious packages may also have been scraped prior to takedown, which could inflate the reported absolute 87.04\% accuracy. Because all three baselines were evaluated under identical experimental conditions, the relative advantage of taint-based slicing remains informative. Future work should use temporally held-out malware datasets to obtain contamination-free estimates.

\section{Conclusion and Future Work}
\label{sec:conclution-future-work}

This paper presents a taint-based code slicing framework for LLM-powered malicious npm package detection. Guided by a curated inventory of JavaScript-specific sensitive APIs, the approach reduces mean input token count by 99.75\% while retaining 73.7\% of malicious features, enabling DeepSeek-Coder-6.7B to achieve 87.04\% detection accuracy and 83.32\% F1 score, outperforming both a naive token-splitting baseline and a CFG-only static slicing approach by nearly 10 percentage points in F1 score. Performance is conditioned on packages invoking the predefined API set; packages that do not use known APIs remain outside the current detection scope.

Several directions remain for future work. First, fine-tuning LLMs on slicing-aware malicious code inputs could reduce false positives and improve robustness against previously unseen attack patterns. Second, deobfuscation preprocessing is a \emph{necessary prerequisite} for production deployment: an LLM-assisted deobfuscator or dynamic sandbox must resolve dynamic call targets into a statically analyzable form before the taint slicer is applied, thereby directly reducing the 44\% empty-slice rate. Third, benchmarking the framework against closed-source LLMs (e.g., GPT-4) will establish an upper performance bound and determine how much of the remaining accuracy gap is attributable to the underlying backbone model versus the input representation. Finally, jointly integrating fine-tuned LLMs, deobfuscation-aware preprocessing, and semantic slicing into an end-to-end pipeline could further strengthen resilience against adversarially crafted packages.

\noindent\textbf{Code \& Data Availability.}
This study uses the malicious package dataset released by Wang et al.~\cite{wang2025malpacdetector} as part of the MalPacDetector project.
All code and replication materials are available at \url{https://github.com/MinhPham131204/replication_package_JSCodeSlicing}

\section*{Acknowledgement}
We acknowledge Ho Chi Minh City University of Technology (HCMUT), VNU-HCM for supporting this study.

\begin{credits}



\end{credits}

{\sloppy\bibliographystyle{splncs04}
\bibliography{ref_shortened}}

\end{document}